\documentclass[12pt]{article}
\usepackage{sc3conf}
\usepackage{amsfonts}
\usepackage{graphicx}

\newcommand{\Sp}{\mathop{\mathrm{span}}}

\newcommand{\bR}{\mathbf{R}}
\newcommand{\cJ}{\mathcal{J}}
\newcommand{\cN}{\mathcal{N}}
\newcommand{\cV}{\mathcal{V}}
\newcommand{\tR}{\tilde{R}}
\newcommand{\rNP}{\mathrm{NP}}

\begin{document}
\raggedbottom

\title{Nonperturbative aspects in $\cN$-fold supersymmetry}

\authors{Toshiaki~Tanaka,\adref{1}  
  and Masatoshi~Sato\adref{2}}

\addresses{\1ad Department of Physics,
 Graduate School of Science,
 Osaka University,\\
 Toyonaka, Osaka 560-0043, Japan,
  \nextaddress \2ad The Institute for Solid State Physics,
 The University of Tokyo,\\
 Kashiwanoha 5-1-5, Kashiwa-shi,
 Chiba 277-8581, Japan.}

\maketitle

\begin{abstract}
Through a nonperturbative analysis on a sextic triple-well potential,
we reveal novel aspects of the dynamical property of the system in
connection with $\cN$-fold supersymmetry and quasi-solvability.
\end{abstract}

\section{Introduction}

\begin{figure}[ht]
\begin{center}
\includegraphics[width=.4\textwidth]{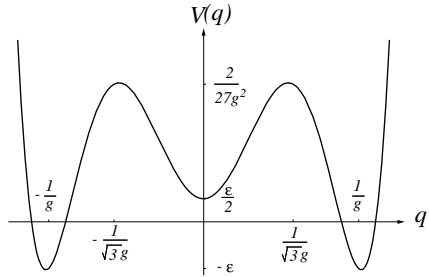}
\caption{The form of the triple-well potential (\ref{eqn:cubpo}).}
\label{fig:tripl}
\end{center}
\end{figure}

Analytical understanding of nonperturbative aspects in quantum theories
is a quite difficult problem.
We will present novel features of the following quantum mechanical
system:
\begin{eqnarray}
H=\frac{p^{2}}{2}+\frac{1}{2}q^{2}(1-g^{2}q^{2})^{2}+\frac{\epsilon}{2}
 (1-3g^{2}q^{2}).
\label{eqn:cubpo}
\end{eqnarray}
This is a triple-well potential having three local minima at $q=0$ and
$q\simeq\pm 1/g$ for $\epsilon g^{2}\ll 1$, see Fig.~\ref{fig:tripl}.
It is a simple one-dimensional quantum mechanical system but the dynamics
and the nonperturbative analysis are highly nontrivial.

We can find that the traditional semi-classical approach fails.
In the case of $\epsilon\neq 0$, there is
a so-called bounce solution as the classical solution which has
a negative mode in the fluctuations. The negative mode contributes
non-zero imaginary part of the spectra in the approximation, showing
instability of the system. Since the spectra of the model must be
real, the instability in the approximation must be \textit{fake}.
In the case of $\epsilon =0$, the classical solution is now
an instanton having no negative mode and does not cause fake instability.
However, another difficulty takes place when we sum up the multi-instanton
contribution with the aid of the dilute-gas approximation. The application
of this approximation leads to the splitting of the ground-state spectra
due to the quantum tunneling, in spite of the fact that the ground-state
of the potential (\ref{eqn:cubpo}) does not degenerate in the tree-level.
As we will see, both the difficulties are circumvented with the aid of
the valley method briefly explained in the next section.
Furthermore, the system (\ref{eqn:cubpo}) possesses significant analytic
properties at specific values of the parameter, namely, $\cN$-fold
supersymmetry and quasi-solvability. These issues are summarized in
Section \ref{sec:7} and \ref{sec:8}.

\section{Valley method}

The main problem in quantum theories concerns with the evaluation
of the Euclidean partition function:
$
Z=\cJ\int\mathcal{D}q\, e^{-S[q]}
$.
Since the evaluation cannot be done exactly in general, one must
find out a proper method which enables one to get a good estimation
of the quantity. The semi-classical approximation is known to be one
of the most established methods. Especially, the uses of instantons
have been succeeded in analyzing non-perturbative aspects of
various quantum systems which have degenerate vacua~\cite{Cole}.
However, validity of the approximation comes into question when
the fluctuations around the classical configuration contain
a negative mode. The appearance of a negative mode indicates
that the classical action does not give the minimum
but rather a saddle point in the functional space.
In this case, one may expect that the path-integral
is dominated by the configurations along the negative mode, which
may intuitively constitute a \textit{valley} in the functional space.
The valley method is a natural realization of this consideration.

At first, we give a geometrical definition of the valley in the
functional space $q(\tau)$~\cite{AK}:
\begin{eqnarray}
\frac{\delta}{\delta q(\tau)}\left[\frac{1}{2}\int\! d\tau'
 \left(\frac{\delta S[q]}{\delta q(\tau')}\right)^{2}
 -\lambda S[q]\right]=0.
\label{eqn:valley1}
\end{eqnarray}
The above definition (\ref{eqn:valley1}) can be interpreted as follows;
for each fixed ``height'' $S[q]$, the valley is defined at the point
where the norm of the gradient vector becomes extremal.
Introducing an auxiliary field $F(q)$, we can make the valley
equation (\ref{eqn:valley1}) a more perspicuous form:
\begin{eqnarray}
\frac{\delta S[q]}{\delta q(\tau)}=F(\tau),\quad
\int\! d\tau'\, D(\tau ,\tau')F(\tau')=\lambda F(\tau).
\label{eqn:valley2}
\end{eqnarray}
where the operator $D$ is defined by
$D(\tau ,\tau')=\delta^{2} S[q]/\delta q(\tau)\delta q(\tau')$.
It is now evident that any solution of the equation of motion is
also a solution of the valley equation (\ref{eqn:valley2}) with
$F(\tau)\equiv 0$.

Next, we separate the integration along the valley line from the
whole functional integration. We parametrize the valley line by
a parameter $R$ and denote the valley configuration by $q_{R}(\tau)$.
Introducing Faddeev-Popov determinant $\Delta[\varphi_{R}]$,
expanding the action $S[q]$ around the valley configuration,
and integrating up to the second order term in the fluctuation
$\varphi_{R}(\tau)=q(\tau)-q_{R}(\tau)$, we finally obtain the
one-loop order result:
\begin{eqnarray}
Z&=&\cJ\int\! dR\int\!\mathcal{D}q\,\delta\left(\int\! d\tau\,
 \varphi_{R}(\tau)G_{R}(\tau)\right)\Delta [\varphi_{R}] e^{-S[q]}
 \nonumber\\
&\simeq&\mathcal{J}\int\frac{dR}{\sqrt{2\pi\det' D_{R}}}
 \Delta [\varphi_{R}]e^{-S[q_{R}]},
\label{eqn:ptfiv}
\end{eqnarray}
where $G_{R}(\tau)$ is the normalized gradient vector on the valley
configuration.
In the above, $\det'$ denotes the determinant in the functional subspace
which is perpendicular to the gradient vector $G_{R}(\tau)$.
The valley equation (\ref{eqn:valley2}) ensure that the subspace
does not contain the eigenvector of the eigenvalue $\lambda$. Therefore,
we can safely perform the Gaussian integrations even when we encounter
a non-positive mode. The extension to the multi-dimensional valley,
which will be needed when there are multiple non-positive eigenvalues,
is straightforward.

\section{Valley-instantons}

Let us investigate the solutions of the valley equation (\ref{eqn:valley2})
for the system (\ref{eqn:cubpo}).
In the case of $\epsilon =0$, the three local minima of
the potential (\ref{eqn:cubpo})
have the same potential value. Thus, there are
(anti-)instanton solutions of the equation of motion which
describe the quantum tunneling between the neighboring vacua:
\begin{eqnarray}
q_{0}^{(I)}(\tau -\tau_{0})=\pm\frac{1}{g}\frac{1}
 {(1+e^{\mp 2(\tau -\tau_{0})})^{1/2}},\quad
q_{0}^{(\bar{I})}(\tau -\tau_{0})=\pm\frac{1}{g}\frac{1}
 {(1+e^{\pm 2(\tau -\tau_{0})})^{1/2}}.
\label{eqn:cubins}
\end{eqnarray}
When $\epsilon\ne 0$, the classical solutions drastically change into
the so-called bounce solutions which cause fake instability. On the
other hand, the solutions of the valley equation (\ref{eqn:valley2})
contain a continuously deformed (anti-)instanton which connects the
two non-degenerate local minima and is called
\textit{(anti-)valley-instanton}~\cite{AKOSW2}.

The solutions of the valley equation (\ref{eqn:valley2}) also
contain a family of the configurations, which tends to the trivial
vacuum configuration in the one limit and tends to well-separated
valley-instanton and anti-valley-instanton configuration in the
other limit. The latter configuration is called $I\bar{I}$-valley.
The bounce solution is also realized as an intermediate configuration
of this family, which is consistent with the fact that the solution
of the equation of motion is also a solution of the valley equations.
For details, see the numerical result in Ref.~\cite{AKOSW2}.
It turns out that the quantum fluctuation around the bounce solution
along this family of the valley configurations actually corresponds
the negative mode. Therefore, we can separate the integration along
the negative mode with the aid of the valley method and thus do not
suffer from the problem of fake instability.

There are two distinct $I\bar{I}$-($\bar{I}I$-)valley configurations
since the curvature at the central potential bottom (at $q=0$) is
different, even at the leading order of $g^{2}$, from the one
at the side potential bottoms (at $q\simeq\pm 1/g$); the $I\bar{I}$-%
($\bar{I}I$-)valley which satisfy $q(\pm T/2)=0$ $(T\gg 1)$ are
different from the ones which satisfy $q(\pm T/2)\simeq 1/g$ or $-1/g$
$(T\gg 1)$.
The Euclidean action of the former with large separation $R$ can be
calculated by the perturbative expansion in
$\lambda\sim O(e^{-2R})$ as follows:
\begin{eqnarray}
S^{(I\bar{I})}(R)=S^{(\bar{I}I)}(R)
 =2S_{0}^{(I)}-\epsilon R+\frac{\epsilon}{2}(T-R)
 -\frac{1}{g^{2}}e^{-2R}+O(e^{-4R}),
\label{eqn:Sibi1c}
\end{eqnarray}
while the one of the latter with large separation $\tR$ can be calculated
in the same way as,
\begin{eqnarray}
S^{(I\bar{I})}(\tR)=S^{(\bar{I}I)}(\tR)
 =2S_{0}^{(I)}+\frac{\epsilon}{2}\tR -\epsilon (T-\tR )
 -\frac{2}{g^{2}}e^{-\tR}+O(e^{-2\tR}),
\label{eqn:Sibi2c}
\end{eqnarray}
where $S_{0}^{(I)}$ denotes the Euclidean action of one (anti-)instanton
Eq.~(\ref{eqn:cubins}) and amounts to $S_{0}^{(I)}=1/4g^{2}$.
In Eqs.~(\ref{eqn:Sibi1c}) and (\ref{eqn:Sibi2c}),
the fourth term can be interpreted as
the interaction term between the valley-instanton and the
anti-valley-instanton. Therefore, the minus sign indicates that
the interaction is attractive.

The other type of the solutions emerges in this case, which is
asymptotically composed of two successive valley-instantons or
two successive anti-valley-instantons. We call them $II$-valley
and $\bar{I}\bar{I}$-valley, respectively. These configurations
do not appear in the case of double-well potentials since they
connect every other vacuum. The Euclidean action
of them with large separation $R$ can be also calculated in
the same way as,
\begin{eqnarray}
S^{(II)}(\tR)=S^{(\bar{I}\bar{I})}(\tR)
 =2S_{0}^{(I)}+\frac{\epsilon}{2}\tR-\epsilon (T-\tR )
 +\frac{2}{g^{2}}e^{-\tR}+O(e^{-2\tR}).
\label{eqn:Siicu}
\end{eqnarray}
Note that the sign of the fourth term is plus and thus the interaction
between the (anti-)valley-instantons in this case is repulsive.
It turns out that these interaction terms in Eqs.~(\ref{eqn:Sibi1c})--%
(\ref{eqn:Siicu}) play an important role in circumventing the breakdown
of the dilute-gas approximation and in producing physically acceptable
results.

\section{Multi-valley-instantons calculus}

\begin{figure}[ht]
\begin{center}
\includegraphics[width=.7\textwidth]{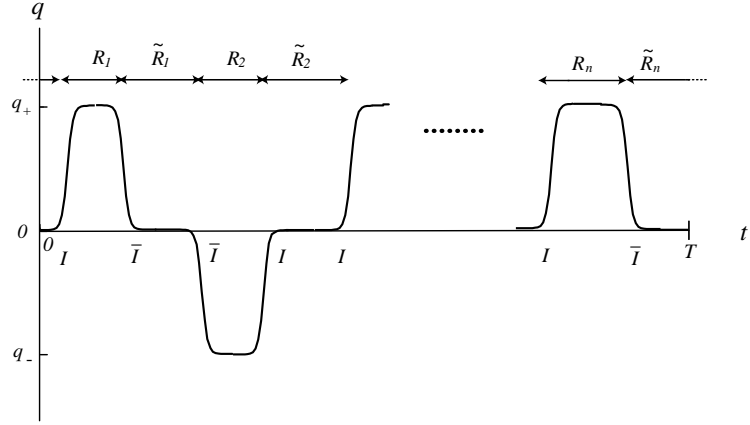}
\caption{The collective coordinates $R_{i}$ and $\tR_{i}$ for a
$2n$ valley-instantons configuration.}
\label{fig:mvccu}
\end{center}
\end{figure}

Utilizing the knowledge of the (anti-)valley-instantons and
the interactions between them obtained previously,
we can evaluate the partition function $Z=\textrm{tr}\, e^{-HT}$
by summing over those configurations made of several
(anti-)valley-instantons.
The sum of the contributions from the $2n$ valley-instantons
configuration can be written as,
\begin{eqnarray}
Z_{\rNP}=\sum_{n=1}^{\infty}\alpha^{2n}
 \sum_{n_{II}=0}^{[n/2]}{n\choose 2n_{II}}\cJ_{n, n_{II}},
\label{eqn:Zcu1}
\end{eqnarray}
where $\alpha^{2}$ denotes the contribution of the Jacobian and
the $R$-independent part of the determinant for one
valley-instanton-pair and is calculated as $\alpha^{2}=\sqrt{2}
e^{-1/2g^{2}}/\pi g^{2}$. The function $\cJ_{n, n_{II}}$ is given by,
\begin{eqnarray}
\lefteqn{
\cJ_{n, n_{II}}=\frac{T}{n}\int_{0}^{\infty}\!
 \left(\prod_{i=1}^{n}dR_{i}\,d\tR_{i}
 \right)\delta\left(\sum_{i=1}^{n}(R_{i}+\tR_{i})-T\right)\exp
 \left[ -( 1-\epsilon )\sum_{i=1}^{n}R_{i}\right.
}\nonumber\\
&&\left. -\frac{1+\epsilon}{2}\sum_{i=1}^{n}
 \tR_{i}+\frac{1}{g^{2}}\sum_{i=1}^{n}e^{-2R_{i}}-\frac{2}{g^{2}}
 \sum_{i=1}^{2n_{II}}e^{-\tR_{i}}+\frac{2}{g^{2}}
 \sum_{i=2n_{II}+1}^{n}e^{-\tR_{i}}\right],
\label{eqn:dfnJcu}
\end{eqnarray}
where $R_{i}$ is the distance between the $(2i-1)$-th and
$2i$-th (anti-)valley-instanton and $\tR_{i}$ the one between
the $2i$-th and the $(2i+1)$-th (anti-)valley-instanton ($\textrm{mod }
n$), see Fig.~\ref{fig:mvccu}.

\section{Nonperturbative shifts of the spectra}\label{sec:5}

From Eqs.~(\ref{eqn:Zcu1}) and (\ref{eqn:dfnJcu}), the nonperturbative
contributions to the spectra are determined by the following equation:
\begin{eqnarray}
\lefteqn{
\alpha^{2}\frac{(-)^{E-\frac{1}{2}-\frac{\epsilon}{2}}\pm 1}{2}
 \left(\frac{2}{g^{2}}
 \right)^{E-\frac{1}{2}-\frac{\epsilon}{2}}\Gamma\left(
 -E+\frac{1}{2}+\frac{\epsilon}{2}\right)\times
}\hspace*{30mm}\nonumber\\
&&\times\left(-\frac{1}{g^{2}}\right)^{\frac{E}{2}-\frac{1}{2}
 +\frac{\epsilon}{2}}\Gamma\left(-\frac{E}{2}+\frac{1}{2}
 -\frac{\epsilon}{2}\right)=1.
\label{eqn:spdcu}
\end{eqnarray}
We will solve the above equation by the series expansion in $\alpha$:
$E_{n_{0}/n_{\pm}}=\sum_{r=0}^{\infty}E_{n_{0}/n_{\pm}}^{(r)}\alpha^{r},$
where $E_{n_{0}}$ stands for the spectra corresponding to, in the
limit $g\to 0$, the eigenfunctions of the center potential well and
$E_{n_{\pm}}$ for the ones corresponding to the parity eigenstates
obtained by the linear combinations of the eigenfunctions of
the each side potential well. The results are complicated because
the calculation needs separate treatments depending on the value of the
parameter $\epsilon$. For details, see Eqs.~(5.22)--(5.27) in
Ref.~\cite{ST1}. From the results, we find that the nonperturbative
contribution to the spectra of the particular levels vanishes at the
specific values of the parameter $\epsilon$ as follows:
\begin{enumerate}
\item $\epsilon =(4\cN +1)/3$ ($\cN =1, 2, 3, \dots$):
 $E_{n_{+}}^{(1)}=E_{n_{+}}^{(2)}=\cdots =0
 \quad\textrm{for}\quad n_{+}<\cN$.
\item $\epsilon =(4\cN -1)/3$ ($\cN =1, 2, 3, \dots$):
 $E_{n_{-}}^{(1)}=E_{n_{-}}^{(2)}=\cdots =0
 \quad\textrm{for}\quad n_{-}<\cN$.
\end{enumerate}

\section{Large-order behavior of the perturbation series}\label{sec:6}

If we evaluate the spectra $E$ by means of the perturbation expansion
in the coupling constant $g^{2}$ as
$E_{n_{0}/n_{\pm}}=E_{n_{0}/n_{\pm}}^{(0)}+\sum_{r=1}^{\infty}
 a_{n_{0}/n_{\pm}}^{(r)}g^{2r}$,
the large order behavior of the coefficients $a^{(r)}$ ($r\gg 1$) can
be estimated via the following dispersion relation~\cite{AKOSW2}:
\begin{eqnarray}
a^{(r)}=-\frac{1}{\pi}\int_{0}^{\infty}\! dg^{2}
 \frac{\textrm{Im}\, E_{\rNP}(g^{2})}{g^{2r+2}}.
\label{eqn:disrel}
\end{eqnarray}
From the results obtained by Eq.~(\ref{eqn:spdcu}),
the leading contributions for $r\gg 1$ reads,
\begin{eqnarray}
a_{n_{0}}^{(r)}&\sim&A_{n_{0}}(\epsilon)\, 2^{r}\,\Gamma\left( r
 +\frac{3}{2}n_{0}+\frac{3}{4}+\frac{3}{4}\epsilon\right),
\label{eqn:lobcu1}\\
a_{n_{\pm}}^{(r)}&\sim&A_{n_{\pm}}(\epsilon)\, 2^{r}\,\Gamma\left(
 r+3n_{\pm}+\frac{3}{2}-\frac{3}{2}\epsilon\right),
\label{eqn:lobcu2}
\end{eqnarray}
where $A_{n}(\epsilon)$'s are some constants depending on $n$ and
$\epsilon$. Equations (\ref{eqn:lobcu1}) and (\ref{eqn:lobcu2}) show
that the perturbative coefficients diverge factorially unless
the prefactor $A_{n}(\epsilon )$'s vanish. It turns out that
the disappearance of the leading divergence takes place only when
$\epsilon =\pm (2n+1)/3$ $(n=1, 2, 3, \dots )$, see Eqs.~(5.31)
in Ref.~\cite{ST1}. More precisely, we obtain the following results:
\begin{enumerate}
\item $\epsilon =(4\cN\pm 1)/3$ ($\cN =1, 2, 3,\ldots$):
 $A_{n_{\pm}}(\epsilon )=0 \quad\textrm{for}\quad n_{\pm}<\cN$.
\item $\epsilon =-(4\cN +1)/3$ ($\cN =1, 2, 3,\dots$):
 $A_{2m_{0}+1}(\epsilon )=0 \quad\textrm{for}\quad m_{0}<\cN$.
\item $\epsilon =-(4\cN -1)/3$ ($\cN =1, 2, 3,\dots$):
 $A_{2m_{0}}(\epsilon )=0 \quad\textrm{for}\quad m_{0}<\cN$.
\end{enumerate}

\section{$\cN$-fold supersymmetry}\label{sec:7}

An $\cN$-fold supersymmetric quantum mechanical system of
one-degree of freedom is, roughly speaking, a system of a pair
of Hamiltonians $H_{\cN}^{\pm}$ which satisfies intertwining
relations with respect to an $\cN$-th order linear differential
operator $P_{\cN}$ as follows:
\begin{eqnarray}
P_{\cN}H_{\cN}^{-}-H_{\cN}^{+}P_{\cN}=0,\quad
 P_{\cN}^{\dagger}H_{\cN}^{+}-H_{\cN}^{-}P_{\cN}^{\dagger}=0.
\label{eqn:inter}
\end{eqnarray}
For the general treatment and discussion on $\cN$-fold supersymmetry,
see Ref.~\cite{AST2}.

It turns out that the system (\ref{eqn:cubpo}) becomes $\cN$-fold
supersymmetric when $\epsilon=\pm (4\cN\pm 1)/3$. More precisely,
when $\epsilon=(4\cN\pm 1)/3$, the relations (\ref{eqn:inter}) are
satisfied for $H_{\cN}^{+}=H$ and following $H_{\cN}^{-}$ and $P_{\cN}$:
\begin{eqnarray}
H_{\cN}^{-}&=&\frac{p^{2}}{2}+\frac{1}{2}q^{2}\left(1-g^{2}q^{2}
 \right)^{2}\nonumber\\
&&{}-\frac{2\cN\mp 1}{6}\left(1-3g^{2}q^{2}\right)+\frac{(2\cN-1\pm 1)
 (2\cN+1\pm 1)}{8q^{2}},\\
P_{\cN}&=&(-i)^{\cN}\prod_{k=-(\cN-1)/2}^{(\cN-1)/2}\left(\frac{d}{dq}
 +q\left(1-g^{2}q^{2}\right)+\frac{\cN\pm 1-2k}{2q}\right).
\end{eqnarray}
When $\epsilon=-(4\cN\pm 1)/3$, the relations (\ref{eqn:inter}) are
satisfied for $H_{\cN}^{-}=H$ and following $H_{\cN}^{+}$ and $P_{\cN}$:
\begin{eqnarray}
H_{\cN}^{+}&=&\frac{p^{2}}{2}+\frac{1}{2}q^{2}\left(1-g^{2}q^{2}
 \right)^{2}\nonumber\\
&&{}+\frac{2\cN\mp 1}{6}\left(1-3g^{2}q^{2}\right)+\frac{(2\cN-1\pm 1)
 (2\cN+1\pm 1)}{8q^{2}},\\
P_{\cN}&=&(-i)^{\cN}\prod_{k=-(\cN-1)/2}^{(\cN-1)/2}\left(\frac{d}{dq}
 +q\left(1-g^{2}q^{2}\right)-\frac{\cN\pm 1+2k}{2q}\right).
\end{eqnarray}

\section{Quasi-solvability}\label{sec:8}

If we define $\cN$-dimensional vector spaces $\cV_{\cN}^{\pm}$ by
$\cV_{\cN}^{-}=\ker P_{\cN}$ and $\cV_{\cN}^{+}=\ker P_{\cN}^{\dagger}$,
we see from Eq.~(\ref{eqn:inter}) that $\cV_{\cN}^{\pm}$ are invariant
under the action of the $\cN$-fold supersymmetric Hamiltonians:
$H_{\cN}^{\pm}\cV_{\cN}^{\pm}\subset\cV_{\cN}^{\pm}$. This means that
both $H_{\cN}^{\pm}$ are quasi-solvable~\cite{quasi}. Therefore,
the system (\ref{eqn:cubpo}) is quasi-solvable when $\epsilon
=\pm (4\cN\pm 1)/3$.

Actually, when $\epsilon=(4\cN\pm 1)/3$,
the solvable sector of $H=H_{\cN}^{+}$ reads,
\begin{eqnarray}
\cV_{\cN}^{+}=\Sp\left\{\phi_{n}^{+}: n=1,\ldots,\cN\right\},
 \; \phi_{n}^{+}(q)=q^{2n-\frac{3}{2}\pm\frac{1}{2}}
 \exp\left(-\frac{g^{2}}{4}q^{4}+\frac{1}{2}q^{2}\right).
\end{eqnarray}
We note that, since $\cV_{\cN}^{+}\subset L^{2}(\bR)$, $\cN$ elements of
the above $\cV_{\cN}^{+}$ are the exact eigenfunctions of the Hamiltonian
(\ref{eqn:cubpo}). They are analytic at $g^{2}=0$ on the $g^{2}$-plane
and thus, (i) the perturbative expansions of them in $g^{2}$ have
nonzero convergent radii, and (ii) there is no nonperturbative
contribution on them. The results (i) and (ii) are also the cases
for the corresponding eigenvalues, which explain the peculiar results
for the case of $\epsilon=(4\cN\pm 1)/3$ in Sections \ref{sec:5}
and \ref{sec:6}, namely, the disappearance of the leading divergence
and the vanishment of the nonperturbative spectral shifts.

Similarly, when $\epsilon=-(4\cN\pm 1)/3$,
the solvable sector of $H=H_{\cN}^{-}$ reads,
\begin{eqnarray}
\cV_{\cN}^{-}=\Sp\left\{\phi_{n}^{-}: n=1,\ldots,\cN\right\},
 \; \phi_{n}^{-}(q)=q^{2n-\frac{3}{2}\pm\frac{1}{2}}
 \exp\left(\frac{g^{2}}{4}q^{4}-\frac{1}{2}q^{2}\right).
\end{eqnarray}
In contrast to in the case of $\epsilon=(4\cN\pm 1)/3$, $\cV_{\cN}^{-}
\not\subset L^{2}(\bR)$ and thus any element of the above $\cV_{\cN}^{-}$
cannot be the exact eigenfunction of the Hamiltonian (\ref{eqn:cubpo}).
However, if we expand them in power of $g^{2}$ as,
$\phi_{n}^{-}(q)=\sum_{r=0}^{\infty}\phi_{n}^{- (r)}(q)g^{2r}$,
they are normalizable at any finite order in $g^{2}$ and thus the elements
of the $\cV_{\cN}^{-}$ give the perturbatively well-defined and correct
eigenfunctions. Therefore, (iii) the perturbative expansions of them in
$g^{2}$ have nonzero convergent radii too, but (iv) there are
nonperturbative contributions on them. The results (iii) and (iv) are also
the cases for the corresponding eigenvalues, which also explain the
peculiar results for the case of $\epsilon=-(4\cN\pm 1)/3$ in Section
\ref{sec:6}, namely, the disappearance of the leading divergence
and the nonvanishment of the nonperturbative spectral shifts.\\


\noindent
\textbf{Acknowledgments.} One of the authors (T.T.) would like to
thank the organizers of the 3rd International Sakharov Conference
on Physics for the invitation. The work by T.T. was supported in
part by a JSPS research fellowship.


\begin{thebibliography}{9}
\newcommand{\J}[4]{\textit{#1} \textbf{#2} (#3) #4}

\bibitem{Cole}
S.~Coleman, in \textit{The Whys of Subnuclear Physics},
 Plenum Publishing Co., New York, (1979).

\bibitem{AK}
H. Aoyama and H. Kikuchi,
 \J{Nucl. Phys.}{B369}{1992}{219}.

\bibitem{AKOSW2}
H. Aoyama, H. Kikuchi, I. Okouchi, M. Sato and S. Wada,
 \J{Nucl. Phys.}{B553}{1999}{644}.

\bibitem{AST2}
H. Aoyama, M. Sato and T. Tanaka,
 \J{Nucl. Phys.}{B619}{2001}{105}.

\bibitem{quasi}
A. G. Ushveridze, \textit{Quasi-exactly Solvable Models in Quantum
 Mechanics}, IOP Publishing, Bristol, (1994), 
 and references cited therein.

\bibitem{ST1}
M. Sato and T. Tanaka,
 \J{J. Math. Phys.}{43}{2002}{3484}.
  
\end{thebibliography}
\end{document}